\documentclass[twocolumn,showpacs,preprintnumbers,amsmath,amssymb,superscriptaddress,prl]{revtex4}

\usepackage[dvips]{graphicx} 
\usepackage{dcolumn}
\usepackage{bm}

\begin{document}

\title{Generalized gauge field theories with non-topological soliton
solutions}

\author{Joaquin Diaz-Alonso}
\affiliation{LUTH, Observatoire de Paris, CNRS, Universit\'e Paris
Diderot. 5 Place Jules Janssen, 92190 Meudon, France}
\affiliation{Departamento de Fisica, Universidad de Oviedo. Avda.
Calvo Sotelo 18, E-33007 Oviedo, Asturias, Spain}
\author{Diego Rubiera-Garcia}
\affiliation{Departamento de Fisica, Universidad de Oviedo. Avda.
Calvo Sotelo 18, E-33007 Oviedo, Asturias, Spain}

\date{\today}

\newcommand{\be }{\begin{equation}}
\newcommand{\bea }{\begin{eqnarray}}
\newcommand{\bh }{\begin{displaymath}}
\newcommand{\en }{\end{equation}}
\newcommand{\ena }{\end{eqnarray}}
\newcommand{\eh }{\end{displaymath}}

\begin{abstract}

We perform a systematic analysis of the conditions under which
\textit{generalized} gauge field theories of compact semisimple
Lie groups exhibit electrostatic spherically symmetric
non-topological soliton solutions in three space dimensions. By
the term \textit{generalized}, we mean that the dynamics of the
concerned fields is governed by lagrangian densities which are
general functions of the quadratic field invariants, leading to
physically consistent models. The analysis defines exhaustively
the class of this kind of lagrangian models supporting those
soliton solutions and leads to methods for their explicit
determination. The necessary and sufficient conditions for the
linear stability of the finite-energy solutions against
charge-preserving perturbations are established, going beyond the
usual Derrick-like criteria, which only provides necessary
conditions.

\end{abstract}

\pacs{05.45.Yv, 11.15.-q, 11.10.Lm, 11.27.+d}

\maketitle

The non-linear Born-Infeld (BI) electrodynamics \cite{born34} is
one of the first consistent relativistic field models found
exhibiting (non-topological) soliton solutions in three space
dimensions (finite-energy and stable electrostatic central-field
solutions, as well as dyon-like solutions and bi-dyon field
configurations \cite{chern99}). In the last two decades there has
been a revival of interest on this model and its extensions to
other fields. These extensions amount to obtain new non-linear
lagrangian models for a given field, by generalizing the (linear
or not) original ones through the same procedure relating the
Maxwell and Born-Infeld lagrangians for electrodynamics (involving
always the well known square root dependence). The models so
obtained have found applications in many different physical
contexts. We mention here, as examples, the generalization of BI
procedure to non-abelian gauge fields in many space dimensions,
suggested by string theory \cite{soliton}, the analysis of
glueballs in terms of soliton solutions of non-abelian BI gauge
field models \cite{g-k00}, the use of these gauge models for the
description of dark energy in the context of Cosmology
\cite{cosmo}, the phenomenological analysis of the nucleon
structure in terms of BI generalization of the quadratic Skyrme
action \cite{skyrme} and the search for self-gravitating gauge
field solutions in General Relativity \cite{gravisoliton}.

However, the choice of Born and Infeld in generalizing the action
of electrodynamics (as well as the above mentioned extensions)
which leads to models supporting stable finite-energy solutions,
is by no means unique. In fact the aim of the present work is to
determine the class of general gauge-invariant lagrangian models
(i.e. defined as arbitrary functions of the field invariants)
which are physically consistent and exhibit this kind of soliton
solutions. This is motivated by their potential utility in the
aforementioned contexts and many others. In \cite{dr07},
\cite{diaz83} we have approached this question for the restricted
cases of (one and many-components) scalar fields
$\phi_{a}(x^{\mu})$, with lagrangian densities defined as
functions $f(X)$ of the kinetic terms $X =
\sum_{a}\partial_{\mu}\phi_{a}.\partial^{\mu}\phi_{a}$. At first
glance this choice seems rather arbitrary from a physical point of
view. Nevertheless it is the natural restriction to the scalar
fields of the general problem considered here for the (abelian and
non-abelian) gauge fields and, as we shall see, those results are
necessary for the present analysis. Let us summarize here the more
useful of those results. We have found in Ref.\cite{dr07} the
general conditions to be satisfied by ``admissible" lagrangian
densities $f(X)$ determining exhaustively the class of models
supporting finite-energy, static spherically symmetric (SSS)
solutions which are linearly stable. We called ``admissible" the
lagrangian densities defined and regular everywhere (class-1
models) or in an open and connected domain including the origin
(class-2 models) and leading in both cases to energy densities
which are positive in these regions and vanish in vacuum. The SSS
solutions obey to first integrals of the field equations which
only contain the field strengths and read \be r^{2}\phi_{a}^{'}(r)
\dot{f}\left(-\sum_{b}\phi_{b}^{'2}(r)\right) = \Lambda_{a},
\label{eq:(0)} \en where $\phi_{a}^{'}(r)=\frac{d\phi_{a}}{dr}$,
$\dot{f}(X)=\frac{df}{dX}$ and the $\Lambda_{a}'s$ are the
integration constants, identified as the scalar charges of the
solutions. The SSS solutions of the many-components scalar fields
may be explicitly written in terms of those of the one-component
scalar models, corresponding to the same functional form of the
lagrangian density, as \be \phi_{a}(r) =
\frac{\Lambda_{a}}{\Lambda} \phi(r,\Lambda) + \chi_{a},
\label{eq:(00)} \en where $\phi(r,\Lambda)$ is the SSS solution of
the one-component scalar field associated with the ``mean-square"
scalar charge $\Lambda = \sqrt{\sum_{a}(\Lambda_{a})^{2}}$, and
the $\chi_{a}'s$ are integration constants. The energy associated
with these solutions, when finite, is degenerate on spheres of
radius $\Lambda$ in the N-dimensional $\Lambda_{a}$ space and
equals the energy of the associated one-component scalar field
with scalar charge $\Lambda$. Moreover, the admissible scalar
field models with soliton solutions have been classified into six
families with regard to the behaviors of the SSS field strengths
at the center ($r=0$) and as $r \rightarrow \infty$. Such families
are the combinations of five characteristic behaviors. Classes A-1
and A-2 correspond to infinite (but integrable) and finite soliton
field strengths at the center, respectively. Classes B-1, B-2 and
B-3 correspond to asymptotic dampings of the soliton field
strengths which can be slower than coulombian, coulombian or
faster than coulombian, respectively. For every class we obtained
the associated behavior of the lagrangian function $f(X)$ around
the corresponding value of X. Finally, all the scalar soliton
solutions of admissible models were shown to be linearly stable
against vanishing scalar charge perturbations.

In the present paper we shall determine the class of generalized
gauge-invariant models supporting similar electrostatic
spherically symmetric (ESS) soliton solutions. Let us begin with
the simpler case of \underline{electromagnetic fields} and go next
to the analysis of non-abelian gauge fields. In the former case we
write the general expression for a relativistic gauge-invariant
lagrangian density as \be L = \varphi(X,Y), \label{eq:(1)} \en
where $\varphi(X,Y)$ is a given function of the two field
invariants \bea && X = -\frac{1}{2}F_{\mu\nu}F^{\mu\nu} =
\vec{E}^{2} - \vec{H}^{2}
\nonumber \\
&& Y = -\frac{1}{2}F_{\mu\nu}F^{*\mu\nu} = 2\vec{E}.\vec{H},
\label{eq:(1bis)} \ena which is assumed to be \textit{continuous
and derivable} in the domain of definition ($\Omega$) of the $X-Y$
plane ($\Re^{2}$). We also assume the condition
$\varphi(X,Y)=\varphi(X,-Y)$ to be satisfied, in order to preserve
parity invariance. As in reference \cite{dr07} we shall call
``class-1" field theories the models for which $\varphi(X,Y)$ is
defined and regular everywhere ($\Omega \equiv \Re^{2}$) and
``class-2" field theories those with $\Omega \subset \Re^{2}$,
$(0,0) \in \Omega$ and $\Omega$ open and connected. Other models
are excluded from this analysis. The original BI electrodynamics
is a class-2 theory corresponding to the particular choice \be
\varphi_{BI}(X,Y) = \frac{1 - \sqrt{1 - \mu^{2}X -
\frac{\mu^{4}}{4}Y^{2}}}{4\pi\mu^{2}}, \label{eq:(2)} \en which
reduces to the Maxwell lagrangian $\varphi(X,Y)=\frac{X}{8\pi}$ in
the low-energy limit ($\mu \rightarrow 0$). We look now for the
conditions to be imposed on the functions $\varphi(X,Y)$ in order
to define consistent general electromagnetic field theories. First
of all let us consider the energy density obtained from the
symmetric (gauge-invariant) energy-momentum tensor: \bea \rho^{s}
&=& T^{s}_{00} = 2\frac{\partial \varphi}{\partial X}\vec{E}^{2} +
2\frac{\partial \varphi}{\partial Y}\vec{E}.\vec{H} - \varphi(X,Y)
\nonumber \\
&=&2X\frac{\partial \varphi}{\partial X} - \varphi(X,Y) +
Y\frac{\partial \varphi}{\partial Y} +
2\frac{\partial\varphi}{\partial X}\vec{H}^{2}. \label{eq:(3)}
\ena The requirements of the positive definiteness of the energy
and the vanishing of the vacuum energy lead to the following
\textit{necessary} conditions to be imposed on $\varphi(X,Y)$ \bea
&& \varphi(0,0) = 0\hspace{0.1cm}; \hspace{0.1cm} \varphi(X,0) < 0
\hspace{0.2cm} (\forall (X < 0,Y = 0) \in \Omega)\hspace{0.1cm};
\nonumber\\
&& \frac{\partial \varphi}{\partial X} \geq 0 \hspace{0.15cm}
(\forall (X,Y) \in \Omega), \label{eq:(4)} \ena aside from the
minimal \textit{necessary and sufficient} condition (to be
satisfied for $\forall (X,Y) \in \Omega$) \bea \rho^{s} \geq
(X+\sqrt{X^{2}+Y^{2}})\frac{\partial \varphi}{\partial X} +
Y\frac{\partial \varphi}{\partial Y} - \varphi(X,Y) \geq 0.
\label{eq:(5)} \ena For the purposes of this study we shall
consider ``admissible" only the models satisfying these
conditions. Then the admissible models are the solutions of the
linear inhomogeneous partial differential equations obtained by
equating the expression in Eq.(\ref{eq:(5)}) to any positive
function and satisfying the conditions (\ref{eq:(4)}).

The field equations obtained from Eq.(\ref{eq:(1)}) take the form
\be
\partial_{\mu}\left(\frac{\partial \varphi}{\partial X}F^{\mu\nu} +
\frac{\partial \varphi}{\partial Y} F^{*\mu\nu}\right) = 0.
\label{eq:(6)} \en For central electrostatic fields, $Y=0$, and
these equations have a first integral, which may be written in
terms of the electrostatic potential $A_{0}(r)$ ($\vec{A}=0,
\vec{E}(r)=-\vec{\nabla}A_{0}(r)$) as \be r^{2}\frac{dA_{0}}{dr}
\frac{\partial \varphi}{\partial X}(X,Y=0) = q, \label{eq:(7)} \en
where $X=(\frac{dA_{0}}{dr})^{2}=E^{2}(r)$ and $q$ is the
integration constant. The solutions of this equation, when
substituted in Eq.(\ref{eq:(6)}), lead to a Dirac $\delta$
distribution of weight $4\pi q$. The definition of the electric
charge associated to a given field \be q = \frac{1}{4\pi}\int
d_{3} \vec{r} \vec{\nabla}.\left(\frac{\partial \varphi}{\partial
X}\vec{E} + \frac{\partial \varphi}{\partial Y} \vec{H}\right),
\label{eq:(8)} \en allows to identify this constant as the
electric charge of the solution \cite{born34}. With the
identification $\phi(r) \equiv A_{0}(r)$, Eq.(\ref{eq:(7)})
coincides with the SSS field equation of a one-component scalar
field model (Eq.(\ref{eq:(0)}) without indices) with a lagrangian
density given by \be L_{scalar} \equiv f(X) = -\varphi(-X,Y=0).
\label{eq:(9)} \en Conversely, we may associate to every
admissible scalar model defined by a lagrangian density $f(X)$, a
family of admissible electromagnetic field models defined by
lagrangian densities $\varphi(X,Y)$ which satisfy
Eqs.(\ref{eq:(9)}) and the conditions (\ref{eq:(4)}) and
(\ref{eq:(5)}). The absolute values of the electrostatic central
field solutions of all these generalizations
($\vert\vec{E}(r,q)\vert$) have the same form, as functions of
$r$, as the SSS \textit{field} solutions ($\phi^{'}(r,\Lambda)$)
of the original scalar model. In calculating the electrostatic
energy, by integration of Eq.(\ref{eq:(3)}), and comparing to the
energy of the associated scalar solitons with the same integration
constant we are lead to the relations \be \varepsilon_{e}(q) =
8\pi q\left[A_{0}(\infty,q) - A_{0}(0,q)\right] -
\varepsilon_{s}(q) = 2\varepsilon_{s}(q), \label{eq:(10)} \en
(indices \textbf{e} and \textbf{s} stand for electric and scalar
fields, respectively). We conclude that the analysis of the
conditions determining the scalar field models with finite-energy
SSS solutions, performed in Ref.\cite{dr07}, allows the
determination of families of admissible electromagnetic field
models supporting finite-energy ESS solutions, and the energy of
such electrostatic solutions of charge $q$ is twofold the energy
of the corresponding scalar soliton with scalar charge $\Lambda =
q$. Equivalently, the scaling of energies in the scalar case
($\varepsilon_s(\Lambda)=\Lambda^{3/2}\varepsilon_s(\Lambda=1)$)
leads to the relation $q=(2)^{2/3} \Lambda$ between the charges of
an electrostatic soliton ($q$) and the associated scalar soliton
($\Lambda$) of equal-energy. Moreover, the classification of the
scalar models through the central and asymptotic behaviors of
their soliton solutions, induces a similar classification of the
gauge models through the behavior of their finite-energy ESS
solutions. The conditions on the lagrangian functions
$\varphi(X,Y)$ leading to the different classes of finite-energy
solutions are immediately obtained, through Eq.(\ref{eq:(9)}),
from those leading to the classification in the scalar case.

However, although in the scalar case the finite-energy SSS
solutions of admissible models are always linearly stable, this is
not so for the generalized gauge-invariant models. These models
must satisfy supplementary conditions in order to support ESS
soliton solutions. To find these conditions let us analyze the
linear stability of the ESS solutions, which requires their energy
to be a minimum against charge-preserving perturbations. Let us
consider a finite-energy ESS solution of the field equations
(\ref{eq:(6)}) ($\vec{E}_{0}(r),\vec{H}_{0}=0$) and introduce a
small regular perturbing field ($\vec{E}_{1}(\vec{r},t)$,
$\vec{H}_{1}(\vec{r},t)$) with vanishing electric charge density.
To the first order, the perturbing fields satisfy the equation \be
\vec{\nabla}.\vec{\Sigma} = 0, \label{eq:(11)} \en where \be
\vec{\Sigma}=\frac{\partial \varphi}{\partial X_{0}}\vec{E}_{1} +
2\frac{\partial^{2} \varphi}{\partial
X_{0}^{2}}(\vec{E}_{0}.\vec{E}_{1})\vec{E}_{0}, \label{eq:(11bis)}
\en and the index $0$ in the derivatives means that they are
calculated for the unperturbed solution. The first-order variation
of the energy functional under these small perturbations, obtained
from the integral of Eq.(\ref{eq:(3)}), becomes \be \Delta_{1}
\varepsilon = -2 \int d_{3}\vec{r}
\vec{\nabla}.\left[A^{0}\vec{\Sigma}\right] + 2 \int d_{3}\vec{r}
A^{0}\vec{\nabla}.\vec{\Sigma}, \label{eq:(12)} \en where we have
introduced the time-like component of the vector potential for the
solution ($\vec{E}_{0} = - \vec{\nabla}A^{0}$). From the
linearized field equations (\ref{eq:(11)}) and the asymptotic
behavior of $\vec{\Sigma}$ (which results from the integration
over all space of the same Eq.(\ref{eq:(11)})), we see that the
first variation of the energy functional vanishes, which is an
extremum condition. The expression of the second-order variation
of the energy functional, obtained from (\ref{eq:(3)}) together
with the expansion of the field equations up to the second order,
reads \bea \Delta_{2} \varepsilon &=& \int d_{3}\vec{r}
\left[\frac{\partial \varphi}{\partial X_{0}}\vec{E}_{1}^{2} +
2\frac{\partial^{2} \varphi}{\partial
X_{0}^{2}}(\vec{E}_{0}.\vec{E}_{1})^{2}\right] +
\nonumber \\
&+&\int d_{3}\vec{r} \left[\frac{\partial \varphi}{\partial
X_{0}}\vec{H}_{1}^{2} - 2\frac{\partial^{2} \varphi}{\partial
Y_{0}^{2}}(\vec{E}_{0}.\vec{H}_{1})^{2}\right]. \label{eq:(12bis)}
\ena By deriving Eq.(\ref{eq:(7)}) with respect to $r$ and taking
into account the monotonic character of $\vert
\vec{E}_{0}(r)\vert$ we can show the positivity of the first
integral for any admissible model. Thus the requirement \be
\frac{\partial \varphi}{\partial X} > 2X\frac{\partial^{2}
\varphi}{\partial Y^{2}}, \label{eq:(13)} \en to be satisfied in
the range of values of $X$ ($Y=0$) defined by the ESS solutions,
is a \textit{necessary and sufficient} condition for the linear
stability of the finite-energy ESS solutions of the
\textit{admissible} models. This stability criterion goes beyond
the widely used Derrick's \textit{necessary} conditions
\cite{derrick64}.

We shall now proceed to extend the analysis to
\underline{non-abelian gauge field} theories of compact semisimple
Lie groups of dimension N. We use the following conventions for
the tensor field strength components in the algebra and their
duals, in terms of the gauge fields $A_{a\mu}$ and the structure
constants $C_{abc}$: \bea
F_{a\mu\nu}&=&\partial_{\mu}A_{a\nu}-\partial_{\nu}A_{a\mu} -
g\sum_{bc} C_{abc} A_{b\mu} A_{c\nu}
\nonumber \\
F^{*}_{a\mu\nu}&=&\frac{1}{2}\epsilon_{\mu\nu\alpha\beta}F_{a}^{\alpha\beta},
\label{eq:(14)} \ena whose components define the fields
$\vec{E}_{a}$, $\vec{H}_{a}$ in the usual way. In building the
lagrangian density functions in this case there is an ambiguity in
the calculation of the traces over the matrix-valued fields
\cite{Tsey77}. We adopt here the ordinary definition of the trace
and introduce the field invariants $X$ and $Y$ as \bea X =
-\frac{1}{2} \sum_{a}(F_{a\mu\nu}F_{a}^{\mu\nu}) =
\sum_{a}\left(\vec{E}_{a}^{2} - \vec{H}_{a}^{2}\right)
\nonumber \\
Y = -\frac{1}{2} \sum_{a}(F_{a\mu\nu}F_{a}^{*\mu\nu}) =
2\sum_{a}\left(\vec{E}_{a}.\vec{H}_{a}\right), \label{eq:(15)}
\ena where $1 \leq a \leq N$. The lagrangian density is now
defined as a given \emph{continuous and derivable} function
$\varphi(X,Y)$ which, for parity invariance, is assumed to be
symmetric in the second argument ($\varphi(X,Y) = \varphi(X,-Y)$).
We also introduce the distinction between class-1 and class-2
models with the same criteria as in the electromagnetic case. The
associated field equations read now \be \sum_{b}D_{ab\mu}
\left[\frac{\partial \varphi}{\partial X}F_{b}^{\mu\nu} +
\frac{\partial \varphi}{\partial Y}F_{b}^{*\mu\nu}\right] = 0,
\label{eq:(16)} \en where $D_{ab\mu} \equiv
\delta_{ab}\partial_{\mu} + g\sum_{c} C_{abc} A_{c\mu}$ is the
gauge covariant derivative. The energy density associated to the
symmetric energy-momentum tensor is \be \rho^{s} = T^{s}_{00} =
2\frac{\partial \varphi}{\partial X}\sum_{a}\vec{E}_{a}^{2} +
2\frac{\partial \varphi}{\partial Y}
\sum_{a}\vec{E}_{a}.\vec{H}_{a} - \varphi(X,Y). \label{eq:(17)}
\en As can be easily verified, the conditions to be imposed on the
lagrangian density for ``admissible" models (positive definiteness
of the energy density and vanishing vacuum energy) take the same
form ((\ref{eq:(4)}), (\ref{eq:(5)})) as in the abelian case.

Let us consider the electrostatic spherically symmetric solutions
of the field equations (\ref{eq:(16)}). We look for fields of the
form \be \vec{E}_{a}(\vec{r}) = -\vec{\nabla}\left(
\phi_{a}(r)\right) =
-\phi^{'}_{a}(r)\frac{\vec{r}}{r}\hspace{0.2cm}; \hspace{0.2cm}
\vec{H}_{a} = 0, \label{eq:(18)} \en where the functions
$\phi_{a}(r) \equiv A^{0}_{a}(r)$ can be identified as the
time-like components of the gauge potential in the Lorentz gauge.
When substituted in the field equations (\ref{eq:(16)}) we are
lead to \bea \vec{\nabla}.\left(\frac{\partial \varphi}{\partial
X}\vec{\nabla}\phi_{a}(r)\right) = 0
\nonumber \\
\sum_{bc} C_{abc}\phi_{b}(r) \phi^{'}_{c}(r) = 0. \label{eq:(19)}
\ena The first group of equations leads to the set of first
integrals \be r^{2}\frac{\partial \varphi}{\partial
X}\phi^{'}_{a}(r) = Q_{a}, \label{eq:(19bis)} \en where
$X=\sum_{c}\phi^{'2}_{c}(r)$ and $Q_{a}$ are the integration
constants, which must be interpreted as the color charges
associated to the soliton. Indeed, as in the abelian case the
color charges associated to a given field read \be Q_{a} =
\frac{1}{4\pi}\int d_{3} \vec{r} \vec{\nabla}.\left(\frac{\partial
\varphi}{\partial X}\vec{E}_{a} + \frac{\partial \varphi}{\partial
Y} \vec{H}_{a}\right), \label{eq:(19ter)} \en which now include
the external source charges and the charges carried by the field
itself. The latter ones vanish for the ESS solutions and it
remains only the former, associated to Dirac distributions of
weight $4\pi Q_{a}$ resulting from the substitution of
Eqs.(\ref{eq:(19bis)}) in the first of (\ref{eq:(19)}).

Equations (\ref{eq:(19bis)}) coincide with the field equations
(\ref{eq:(0)}) for a multicomponent SSS scalar field, whose
lagrangian density is given by \be L =
f\left(\sum_{a}\partial_{\mu}\phi_{a}.\partial^{\mu}\phi_{a}\right)
\equiv f(X) = -\varphi(-X,Y=0). \label{eq:(20)} \en Consequently
their solutions coincide with those of Eq.(\ref{eq:(0)}) for the
associated N-components scalar field theory, and reduce finally to
the solution of the associated one-component scalar problem. These
solutions are then given by Eq.(\ref{eq:(00)}) with the
substitution $\Lambda_{a} \rightarrow Q_{a}$. They must also
satisfy the second set of equations (\ref{eq:(19)}). This
restricts the possible values of the constants $\chi_{a}$ and
leads to the following final expression of the general ESS
solutions \be \phi_{a}(r) = \frac{Q_{a}}{Q}\left(\phi(r,Q) +
\chi\right), \label{eq:(21)} \en where $Q =
\sqrt{\sum_{a}(Q_{a}^{2})}$ is the ``mean-square" color charge and
$\chi$ is an additive constant which is the same for all the
components of the potential. In terms of the fields these
solutions read \be \vec{E}_{a}(r) =
-\frac{Q_{a}}{Q}\phi^{'}(r,Q)\frac{\vec{r}}{r} \hspace{0.2cm};
\hspace{0.2cm} \vec{H}_{a} = 0. \label{eq:(22)} \en The associated
energy is obtained from the integration of Eq.(\ref{eq:(17)}). It
is now straightforward to show that (as in the many-components
scalar case) this energy, if finite, is degenerate on spheres of
radius $Q$ in the color charge space and is related to the energy
of the corresponding scalar soliton of charge $\Lambda = Q$
through the same equation (\ref{eq:(10)}) of the abelian case.
Moreover, the criteria obtained in Ref.\cite{dr07} determining the
classes of admissible scalar field models with SSS soliton
solutions determine also, through Eq.(\ref{eq:(20)}), the classes
of admissible generalized non-abelian gauge models supporting
finite-energy electrostatic particle-like solutions.

As in the electromagnetic case we must look now for the conditions
leading to the stability of these ESS solutions. Nevertheless, as
already mentioned, a new qualitative effect arises in the present
situation, which is related to the intrinsic color charge density
associated to the fields perturbing the ESS solutions. By defining
$E_{1a}(\vec{r},t)$, $H_{1a}(\vec{r},t)$ and
$A_{1a}^{\mu}(\vec{r},t)$ as the perturbing fields and potentials
respectively, we must require that the total color charges of the
perturbed solution remain unchanged. To lowest order the first of
Eqs.(\ref{eq:(19)}) leads to \be \vec{\nabla}.\vec{\Sigma}_{a} =
\Delta_{1} n_{a}, \label{eq:(23)} \en where the corrections to the
color charge density distributions introduced by the
perturbations, $\Delta_{1} n_{a}(\vec{r},t)$, and the term
$\vec{\Sigma_{a}}$ are given by \bea \Delta_{1} n_{a} &=& -
g\frac{\partial \varphi}{\partial X_{0}} \sum_{bc} C_{abc}
\vec{A}_{1b}.\vec{E}_{c}
\nonumber \\
\vec{\Sigma}_{a} &=& \frac{\partial \varphi}{\partial
X_{0}}\vec{E}_{1a} + 2\frac{\partial^{2} \varphi}{\partial
X_{0}^2} \sum_{b}(\vec{E}_{b}.\vec{E}_{1b})\vec{E}_{a}.
\label{eq:(24)} \ena These corrections must satisfy the
charge-preserving condition $\frac{1}{4\pi}\int d_{3}\vec{r}
\Delta_{1} n_{a} = 0$, which determines the asymptotic behavior of
$\vec{\Sigma_{a}}$ through the integration of Eq. (\ref{eq:(23)}).
The lowest-order perturbation of the energy, obtained from the
integral of (\ref{eq:(17)}), using (\ref{eq:(23)}) and this
charge-preserving condition becomes \be \Delta_{1}\varepsilon =
2\int d_{3}\vec{r} \sum_{a} \phi_{a} \Delta_{1} n_{a},
\label{eq:(25)} \en which must be interpreted as the interaction
energy of the field of the solution with the first-order
correction to the charge density distribution introduced by the
perturbations. This contribution vanishes identically due to the
second set of equations (\ref{eq:(19)}). Consequently, the first
variation (\ref{eq:(25)}) vanishes, and the solution is an
extremum of the energy functional.

In the same way we obtain the second variation of the energy by
perturbing up to the second order the energy functional and the
field equations. The final expression is \bea \Delta_{2}
\varepsilon &=& \int d_{3}\vec{r} \left[\frac{\partial
\varphi}{\partial X_{0}} \sum_{a}\vec{E}_{1a}^{2} +
2\frac{\partial^{2} \varphi}{\partial X_{0}^{2}}
\left(\sum_{a}\vec{E}_{0a}.\vec{E}_{1a}\right)^{2}\right] +
\nonumber \\
&+&\int d_{3}\vec{r} \left[\frac{\partial \varphi}{\partial X_{0}}
\sum_{a}\vec{H}_{1a}^{2} - 2\frac{\partial^{2} \varphi}{\partial
Y_{0}^{2}}
\left(\sum_{a}\vec{E}_{0a}.\vec{H}_{1a}\right)^{2}\right]
\label{eq:(26)} \ena The first integral is always positive. Thus
the requirement of positivity of $\Delta_{2} \varepsilon$ leads to
the condition \be \frac{\partial \varphi}{\partial X} >
2X\frac{\partial^{2} \varphi}{\partial Y^{2}}, \label{eq:(27)} \en
to be satisfied in the domain of values of $X$ $(Y=0)$ defined by
the solutions. This condition, which is formally the same as in
the abelian case, is \textit{necessary and sufficient} for the
linear stability of finite-energy ESS solutions of admissible
non-abelian gauge models.

As already discussed in Ref.\cite{dr07}, the \textit{linear}
stability of these non-topological solitons does not guarantee
their \textit{general} stability, intended as the preservation of
the ``soliton identity" when interacting with strong ``external"
fields. Only in some few special models with \textit{topological}
soliton solutions (mainly in one-space dimension) the existence of
conserved topological charges allows the ``detection" of the
presence of solitons in a field configuration \cite{scott73}. But
despite some tentative approaches to this question \cite{chern98},
the development of general methods for the analysis of the
stability and dynamical evolution of strongly-interacting solitons
in three space dimensions remains an unsolved problem.
Nevertheless, for low-energy interactions (weak external fields),
the identity of the soliton is preserved if (and only if) linear
stability holds. In this case an expansion in the characteristic
energy scale of interaction (or, equivalently, in the
characteristic value of the intensity of the external field)
becomes possible. At the leading order this expansion allows the
interpretation of the field dynamics in terms of particle-field
force laws and radiative behavior \cite{chern99}, \cite{chern98}.

Let us conclude with a brief comment about the important question
of the radiative solutions of the generalized gauge field
theories. In this work we have focused on the static soliton
solutions of these theories and, from this point of view, the
Born-Infeld-like models are only particular examples of a large
family, without special distinguishing properties. However, the BI
original abelian version has very special properties from the
point of view of wave propagation. Indeed, as can be easily
verified, all these theories exhibit plane wave solutions
propagating with the speed of light but, owing to the
non-linearity, there are also other radiative solutions
propagating with more complex dispersion relations. In most cases
the energy of such waves tends to cumulate, evolving towards
spatially-singular configurations, with formation of shocks after
some critical time. A system of field equations whose wave
solutions are free from this evolution is called ``completely
exceptional". In analyzing this question for the case of
generalized electromagnetic field theories, G. Boillat
\cite{boillat70} established that, from among the family of all
admissible models belonging to the class B-2 defined here
(asymptotic coulombian behavior of the ESS solutions; the
condition $\frac{\partial \varphi}{\partial X}(X=0,Y=0) = 1$
imposed on the lagrangian excludes the classes B-1 and B-3 from
the Boillat analysis) only the Born-Infeld model is completely
exceptional. However, the analysis of modern versions of the
Born-Infeld model, in the contexts of Kaluza-Klein or string
theories, concludes that this exceptionality is lost in the
generalizations of the classical version \cite{kern99}.

In summary, by extending the methods developed in Ref.\cite{dr07}
for the scalar field models, the present analysis allows the
explicit determination of any admissible generalized gauge field
model supporting ESS soliton solutions. More details, examples and
applications of these results will be published soon \cite{dr071}.
\\

\textbf{Acknowledgements}\\

We are grateful to Drs. E. Alvarez, B. Carter and V. Vento for reading the
manuscript and encouraging comments.

\end{document}